\documentclass[preprint2]{emulateapj}

\newcommand{\ie}{{\it i.e.~}}
\newcommand{\vs}{{\it vs.~}}
\newcommand{\eg}{{\it e.g.~}}

\newcommand{\fig}{\textsc{Fig.~}}
\newcommand{\bA}{\textit{\textbf{A}}}
\newcommand{\bB}{\textit{\textbf{B}}}

\newcommand{\FLor}{{\textit{\textbf{F}}_{\!{\mathcal{L}}}}}

\newcommand{\er}{\mathbf{\hat{e}}_{r}}
\newcommand{\etheta}{\mathbf{\hat{e}_{\theta}}}
\newcommand{\ephi}{\mathbf{\hat{e}_{\varphi}}}

\usepackage{graphicx}	
\usepackage{dcolumn}	
\usepackage{bm}		
\usepackage{natbib}
\begin{document}

\title{
On the Stability of Non Force-Free Magnetic Equilibria in Stars
}


\author{V. Duez, J. Braithwaite}
\affil{Argelander Institut f\"ur Astronomie, Universit\"at Bonn, Auf dem H\"ugel 71, D-53111 Bonn, Germany}
\email{vduez@astro.uni-bonn.de}
\and
\author{S. Mathis}
\affil{Laboratoire AIM, CEA/DSM-CNRS-Universit\'e Paris Diderot, IRFU/SAp Centre de Saclay, F-91191 Gif-sur-Yvette, France}

\begin{abstract}
The existence of stable magnetic configurations in white dwarfs, neutron stars and various non-convective stellar {regions} is now well recognized. 
It has recently been shown numerically that various families of equilibria, including axisymmetric mixed poloidal-toroidal configurations, are stable. 
Here we test the stability of an analytically-derived non force-free magnetic equilibrium, using three-dimensional magnetohydrodynamic simulations: the mixed configuration is compared with the dynamical evolution of its purely poloidal and purely toroidal components, both known to be unstable. 
The mixed equilibrium shows no sign of instability under white noise perturbations. 
{This configuration therefore provides a good description of magnetic equilibrium topology inside non-convective stellar objects and will be useful to initialize magneto-rotational transport in stellar evolution codes.}
\end{abstract} 
%
\maketitle
%
\section{\label{intro} Introduction}
Magnetic fields are being detected more and more routinely at the surface of many stars, and are responsible for various physical phenomena likely to deeply modify our traditional vision of stellar evolution, especially during their early and late stages. The presence of even a relatively weak magnetic field can have an important impact on the collapse and fragmentation of prestellar cores \citep{Commercon:2010}, as well as influencing the rotation rate of {pre-main-sequence} stars \cite[see e.g.][]{Alecian:2008}. On the other side of the {Hertzsprung-Russell} diagram its feedback effects may also play a key role in supernovae and mechanical energy deposition in the {interstellar medium}, for instance. \\
Magnetic fields are also an important actor {in main-sequence stars}. First, we cannot dismiss the possibility of a large-scale magnetic field being responsible for the quasi-uniform rotation behaviour in the bulk of the solar radiation zone, as revealed by $p$-modes helioseismology \citep[][]{Eff-Darwich:2008}.
Second, strong fields ($300$ G to $30$ kG) are observed via the Zeeman effect in some fraction of main-sequence A stars \citep[the Ap stars, see][]{Auriere:2007}, as well as {in} B stars and {in} a handful of O stars \citep{Grunhut:2009}. The bimodality of rotational {velocities} observed among Ap \vs normal A stars shows the critical effect of magnetic fields on rotation and therefore also on meridional circulation and chemical transport \citep[see][]{Mathis:2005}. Finally, magnetic white dwarfs display fields strength of $10^4 - 10^9$ G, and neutron stars host fields in the range $10^8-10^{15}$ G, in  both cases detected using several distinct methods. \\
The large-scale, ordered nature of these fields (often approximately dipolar) and the scaling of their strengths as a function of their host properties (according to the flux conservation scenario) favour a fossil hypothesis, whose origin is not yet elucidated. \\

Another fundamental question is the topology of these large-scale magnetic fields. To have survived since the star's formation, a field must be stable on a dynamic (Alfv\'en) timescale. It was suggested by \cite{Prendergast:1956} that a stellar magnetic field in stable axisymmetric equilibrium must contain both poloidal (meridional) and toroidal (azimuthal) components, since both are unstable on their own \citep{Tayler:1973,Wright:1973,Braithwaite:2006b,Bonanno:2008}.
This was confirmed recently by numerical simulations by \cite{Braithwaite:2004, Braithwaite:2006a} who showed that an arbitrary initial field evolves on an Alfv\'en timescale into a stable configuration; axisymmetric mixed poloidal-toroidal fields were found. {Once formed, it continues to evolve on longer timescales through diffusive processes such as finite conductivity: the field then moves outwards, passing quasi-statically through a series of stable axisymmetric equilibria until it changes eventually to a non-axisymmetric equilibrium. These non-axisymmetric equilibria are described in more detail in \cite{Braithwaite:2009}. }
%
\section{The relaxed non force-free configuration}
\label{analytic}
Here, we deal with axisymmetric, non force-free magnetic configurations {(\ie with a non-zero Lorentz force)} in equilibrium inside a conductive fluid. We first restrict ourselves to the non-rotating case, but results also apply to rotating stars where rotation is uniform \citep{Woltjer:1959b}, which could be the case if magnetic field is strong enough, and where meridional circulation can be neglected \citep[i.e. when the star does not loose angular momentum and have a stationary structure: see][]{Busse:1981,Zahn:1992,Decressin:2009}. The more general case including {(differential) rotation (and induced meridional circulation)} will be studied in a forthcoming work. 
Several reasons inclined us to focus on non force-free equilibria instead of force-free ones; let us briefly describe them here. 
First, \citet{Reisenegger:2009} reminds us that no configuration can be force-free everywhere. Although there do exist ``force-free" configurations, they must be confined by some region or boundary layer with non-zero or singular Lorentz force. Discontinuities such as current sheets are unlikely to appear in nature except in a transient manner.
Second, non force-free equilibria have been identified in plasma physics as the result of relaxation {(self-organization process involving magnetic reconnections, in resistive MHD)}, \eg by \citet{Montgomery:1988,Shaikh:2008}.
Third, as shown by \cite{Duez:2010b}, this family of equilibria is a generalization of Taylor states {(force-free relaxed equilibria)} in a stellar context, where the stratification of the medium plays a crucial role.
%
\subsection{The magnetic field in MHS equilibrium}
Let us briefly recall the assumptions made in building the semi-analytical model of magnetohydrostatic (MHS) equilibrium described by \cite{Duez:2010b}.
The axisymmetric magnetic field $\bB(r, \theta)$ is expressed as a function of a poloidal flux $\Psi(r,\theta)$, a toroidal potential $F(r,\theta)$, and the potential vector $\bA\left(r,\theta\right)$ so that it is divergence-free by construction:
\begin{eqnarray}
\bB =
 \frac {1}{r \sin \theta} \left( {\bm\nabla} \Psi \:\mathbf{\times} \:\ephi + F \;\ephi \right)={{\bm\nabla}}\times{\bA},
\label{bofpsiandf}
\label{BPsiF}
\end{eqnarray}
where in spherical coordinates the poloidal component is in the meridional plane ($\er,\etheta$) and the toroidal component is along the azimuthal direction ($\ephi$).
The MHS equation expressing balance between the pressure gradient force, gravity and the Lorentz force is 
\begin{equation}
{\textit{\textbf{0}}} = -{\bm\nabla}\: P-\rho\:{\bm\nabla}\: V+\frac{1}{\mu_0}({\bm\nabla}\times{\bB})\times{\bB},
\label{MHS1}
\end{equation}
where $V$ is the gravitational potential which satisfies the Poisson equation : 
$\nabla^{2}V=4\pi G\rho$.
%
\subsection{The non force-free barotropic equilibrium family}
%
\subsubsection{A variational approach}
Here, we focus on the minimum energy non force-free MHS equilibrium in stably stratified radiation zones. Given the field strengths in real stars, the ratio of the Lorentz force to gravity is very low: stellar interiors are in a regime where $\beta=P/P_{\rm Mag}>\!\!>1$, $P_{\rm Mag}=B^2/\left(2\mu_0\right)$ being the magnetic pressure. We then identify the two MHD invariants {governing the evolution of the reconnection phase, that leads to} relaxed states in the non force-free context: the magnetic helicity ${\mathcal H}=\int_{\mathcal V}{\bA}\cdot{\bB}\:{\rm d}{\mathcal V}$ and the mass encompassed in poloidal magnetic surfaces $M_{\Psi}=\int_{\mathcal V}\Psi\:{\overline \rho}\:{\rm d}{\mathcal V}$, conserved because of the stable stratification. Assuming a selective decay during relaxation (the magnetic energy decays much faster than ${\mathcal H}$ and $M_{\Psi}$ {so that they can be considered constant on an energetic decay e-folding time}), a variational method allows us to derive the elliptic linear partial differential equation governing $\Psi$ \citep{Woltjer:1959b,Duez:2010b}:
\begin{equation}
\Delta^{*}\Psi +\frac{\lambda_1^2}{R^2}\, \Psi = - \mu_0\,\overline\rho\, r^2 \sin^2 \theta\,\beta_0.
\label{GSDM}
\end{equation}
{Here,} $\overline \rho$ is the density in the non-magnetic case, $\Delta^* \Psi \equiv \partial_{rr}{\Psi} + \sin \theta \: \partial_{\theta} \left(\partial_{\theta}{\Psi}/\sin \theta\right)/r^2$ the Grad-Shafranov operator in spherical coordinates, $\lambda_1$ a coefficient to be determined, $R$ a characteristic radius, and $\beta_0$ is constrained by the field's intensity. This equation is similar to the Grad-Shafranov equation used to find equilibria in magnetically confined plasmas \citep{Grad:1958, Shafranov:1966}, the source term being here related to the stellar structure through $\overline{\rho}$ \citep[see][for a discussion of the general form of this equation in astrophysics]{Heinemann:1978}. Furthermore, this equilibrium is in a barotropic state
(in the {\it hydrodynamic} meaning of the term, \ie isobar and iso-density surfaces coincide) where the field is explicitly coupled with stellar structure through: ${\bm\nabla}\times\left(\FLor/{\overline\rho}\right)={\textit{\textbf{0}}}$, where $\FLor$ is the Lorentz force. This is a generalization of Prendergast's equilibrium taking into account compressibility, first studied in polytropic cases by \cite{Woltjer:1960}.
%
\subsubsection{Solution}
\begin{figure}[h!]
\begin{center}
\includegraphics[width=0.475\textwidth]{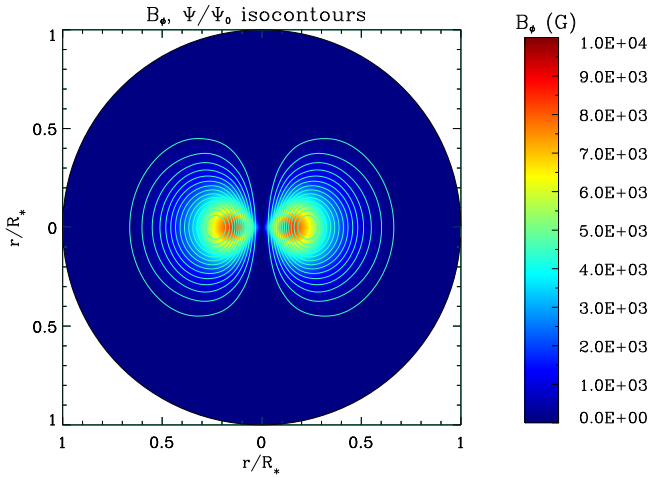}
\end{center}
\caption{Toroidal magnetic field strength in colorscale (arbitrary field's strength) and normalized isocontours of the poloidal flux function $\Psi$ in meridional cut 
for the first equilibrium configuration ($\lambda_{1}^{1}\simeq33$). The neutral line is located at $r\simeq0.23\:R_{*}$. 
\label{FigBPsi}}
\end{figure}

The boundary conditions have now to be discussed. 
In \cite{Duez:2010a, Duez:2010b}, we considered the general case of a field confined between two radii, owing to the possible presence of both a convective core and a convective envelope and to ensure the conservation of magnetic helicity. 
We here choose to cancel both radial and latitudinal fields at the surface, to avoid any current sheets, conserving once again magnetic helicity. Owing to its small extension, the {possible effects of the} convective core on the large-scale surrounding field are neglected.
Using Green's function method we finally obtain the purely dipolar, general solutions indexed by $i$:
\begin{eqnarray}
\Psi_i\left(r,\theta \right)
&=&
-
\mu_0\beta_0\frac{\lambda_{1}^{i}}{R}r \Bigg\{j_{1}\left(\lambda_{1}^{i}\frac{r}{R}\right)\int_{r}^{R}\!\left[y_{1}\left(\lambda_{1}^{i}\,\frac{\xi}{R}\right)\overline\rho \xi^3\right]\!{\rm d}\xi
\nonumber\\
&+&
 y_{1}\left(\lambda_{1}^{i} \frac{r}{R}\right)\!\int_{0}^{r}\left[
j_{1}\left(\lambda_{1}^{i}\frac{\xi}{R}\right)\overline\rho\xi^3\!\right]\!{\rm d}\xi
\Bigg\}\: \sin^2 \theta,\quad\;
\end{eqnarray}
$R$ being the upper boundary confining the magnetic field; $\lambda_{1}^{i}$ are the set of eigenvalues indexed by $i$ allowing to verify the boundary conditions.
{The functions} $j_{l}$ and $y_{l}$ are respectively the spherical Bessel functions of the first and the second kind.  {As shown in \cite{Duez:2010b}, the first radial mode is the lowest energy state. We thus focus here only on this mode $i=1$}.
\noindent The toroidal magnetic field is then given using $F(\Psi) = \lambda_1 \Psi/R$; in the case of a {\it stably stratified} $n=3$ polytrope and for the simulation purposes where we set $R=0.85\: R_*$, we have $\lambda_1\simeq 32.95$, while for a constant density profile (in a zero gravity medium), we have $\lambda_1 \simeq 5.76$. 
The solution for the n=3 polytrope is represented in \fig \ref{FigBPsi}. 
\begin{figure}[h!]
\begin{center}
\includegraphics[width=0.475\textwidth]{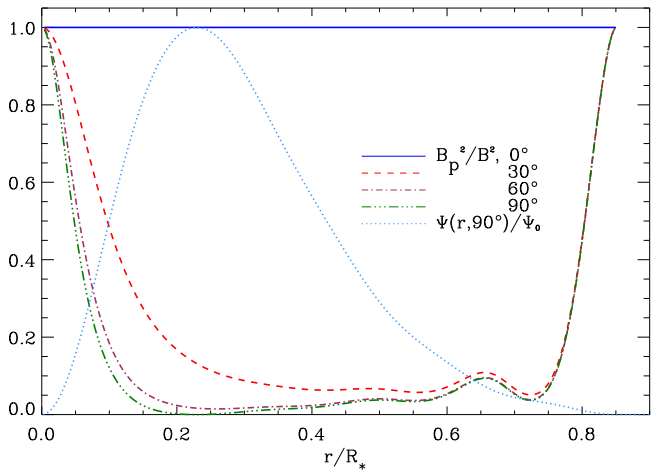}
\end{center}
\caption{Poloidal to total magnetic energy density $B^2_{\rm p}/B^2$ as a function of the radius for different colatitudes, and radial profile of the poloidal flux function $\Psi$ at the equator. 
\label{FigEp_E}}
\end{figure}
{The ratio of the poloidal to total magnetic energy density $B^2_{\rm p}/B^2\:(r,\theta)$ is plotted as a function of the radius in \fig \ref{FigEp_E}, for various latitudes. Notice that its integrated value $E_{\rm p}/E \simeq 5.23 \times 10^{-2}$ is in the range of ratios found in stable axisymmetric equilibria forming in simulations from random small-scale initial conditions \citep{Braithwaite:2006a}.}
%
\section{Stability: numerical method}
%
\subsection{The numerical model} 
The setup of the numerical model is similar to that in Braithwaite \& Nordlund 2006, where a fuller account can be found; a brief outline is given here.
We use the {\textsc{Stagger} code} \citep{Nordlund:1995}, a high-order finite-difference Cartesian MHD code containing a ``hyper-diffusion'' scheme. We use a resolution of $192^3$.

We model the star as a self-gravitating ball of ideal gas ($\gamma=5/3$) with radial density and pressure profiles initially obeying the polytropic (thus barotropic) relation $P \propto \rho^{1+(1/n)}$, with index $n=3$ -- a good approximation to an upper-main-sequence star. 
It seems unlikely that a different EOS, for instance that of a neutron star, will make even much quantitative difference to the results. 
The important point is the stable stratification. 

We use this model to compare the dynamical evolution of the mixed (poloidal-toroidal) configuration to that of its purely poloidal and toroidal components on their own, both of which are unstable as mentioned above.
We should therefore see these instabilities, growing on an Alfv\'en timescale. 
{To test the stability of the configurations, we add a random ``white noise'' perturbation to the density field. The perturbation in density (1\% in amplitude) contains length scales ranging from $R_\ast$ to $0.08R_\ast$, the latter being double the Nyquist wavelength. This is roughly equivalent to azimuthal wavenumbers up to $m=38$ at a radius of $R_\ast/2$. }
%
\subsection{Results}

\begin{figure*}[!htp]
\begin{center}
\includegraphics[width=0.32\hsize,angle=0]{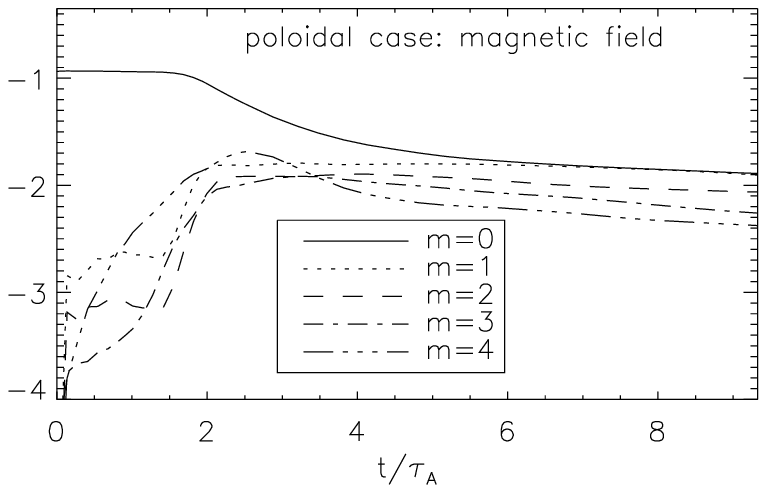}
\includegraphics[width=0.32\hsize,angle=0]{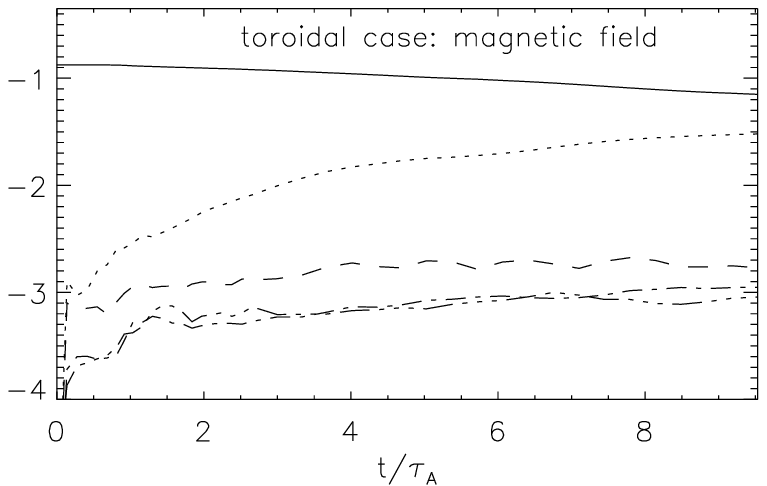}
\includegraphics[width=0.32\hsize,angle=0]{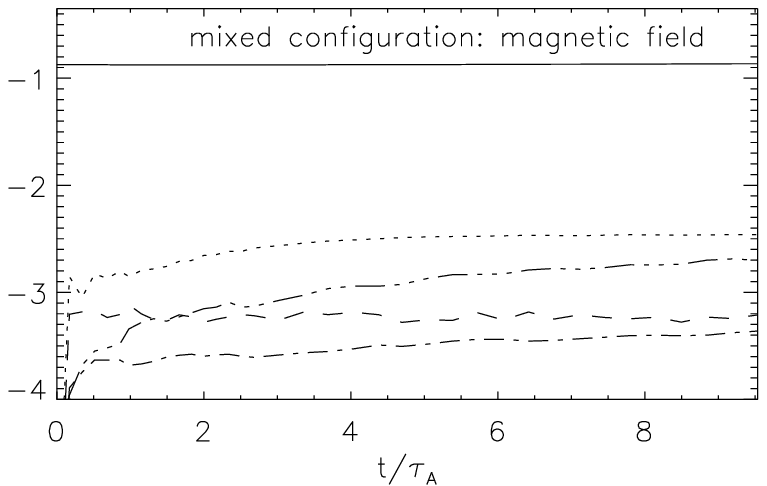}\\
\includegraphics[width=0.32\hsize,angle=0]{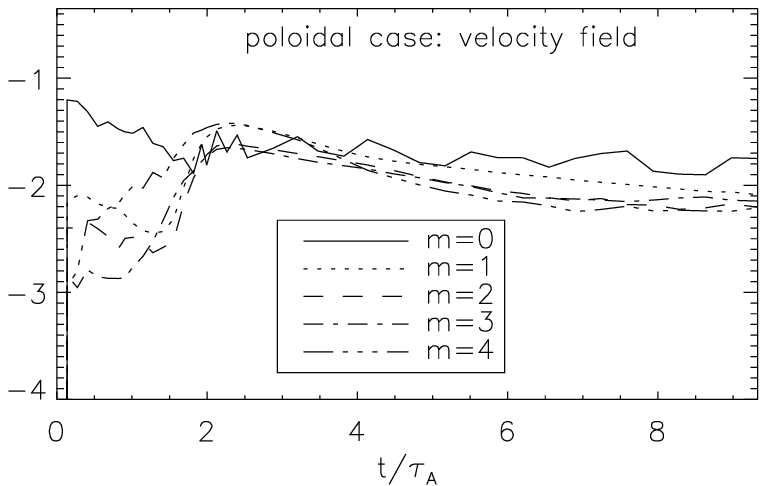}
\includegraphics[width=0.32\hsize,angle=0]{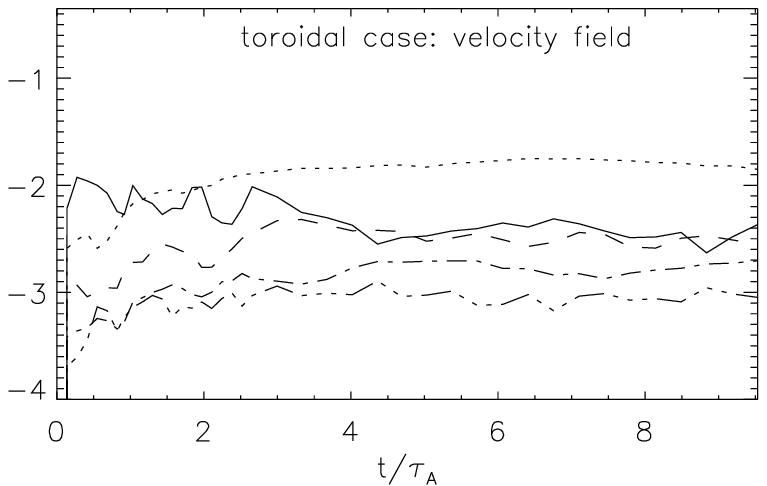}
\includegraphics[width=0.32\hsize,angle=0]{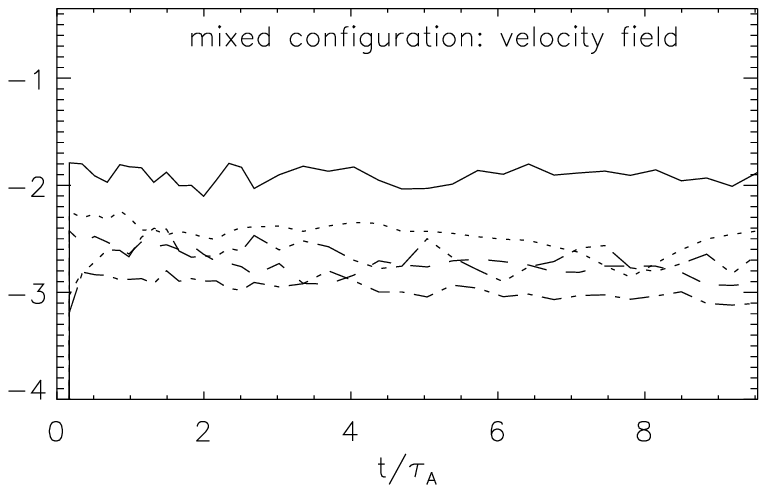}
\end{center}
\caption{Time evolution of the (log) amplitudes in azimuthal modes $m=0$ to $4$ averaged over the stellar volume of the magnetic field (top row) and the velocity field (bottom row) in the simulations with the purely poloidal field (left), purely toroidal field (middle) and the mixed field (right). Initially, all the magnetic energy is in the $m=0$ mode since the initial conditions are axisymmetric. }
\label{fig:mmodes}
\end{figure*}

\begin{figure}[!htp]
\parbox[t]{0.237\textwidth}{\caption{\label{fig:vapor}
Magnetic field lines representing the mixed field configuration (right), looking along the axis (top) and from the side (middle and bottom). The purely toroidal component of the field is represented on the left. {The colorscale is a function of the density.} {\it Upper and middle panels :} configurations at $t=0$; {\it Lower panels :} configuration at $t=10\: \tau_A$.}}
\parbox[t]{0.237\textwidth}{\mbox{}\\\includegraphics[width=0.235\textwidth,angle=0]{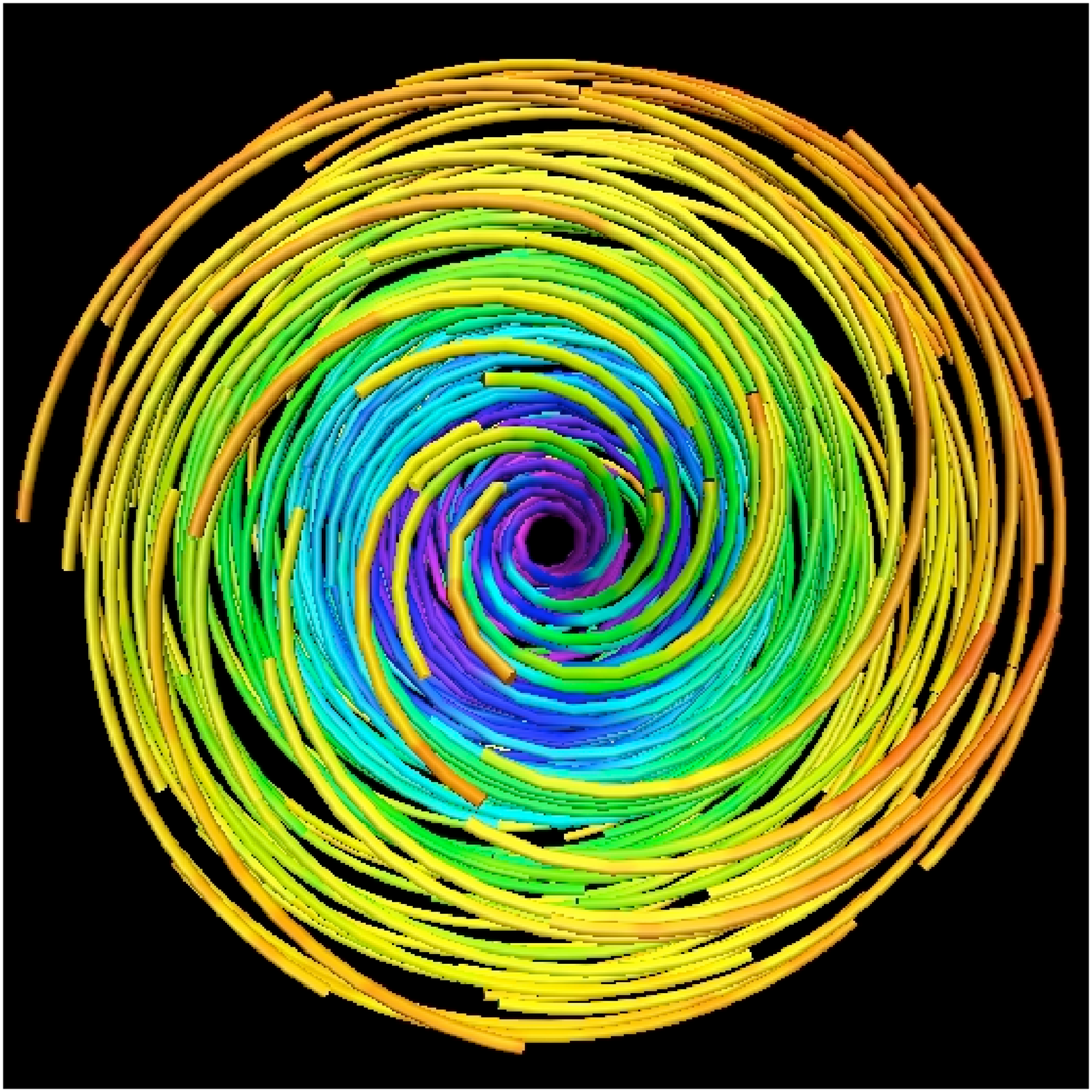}}\\
\includegraphics[width=0.235\textwidth,angle=0]{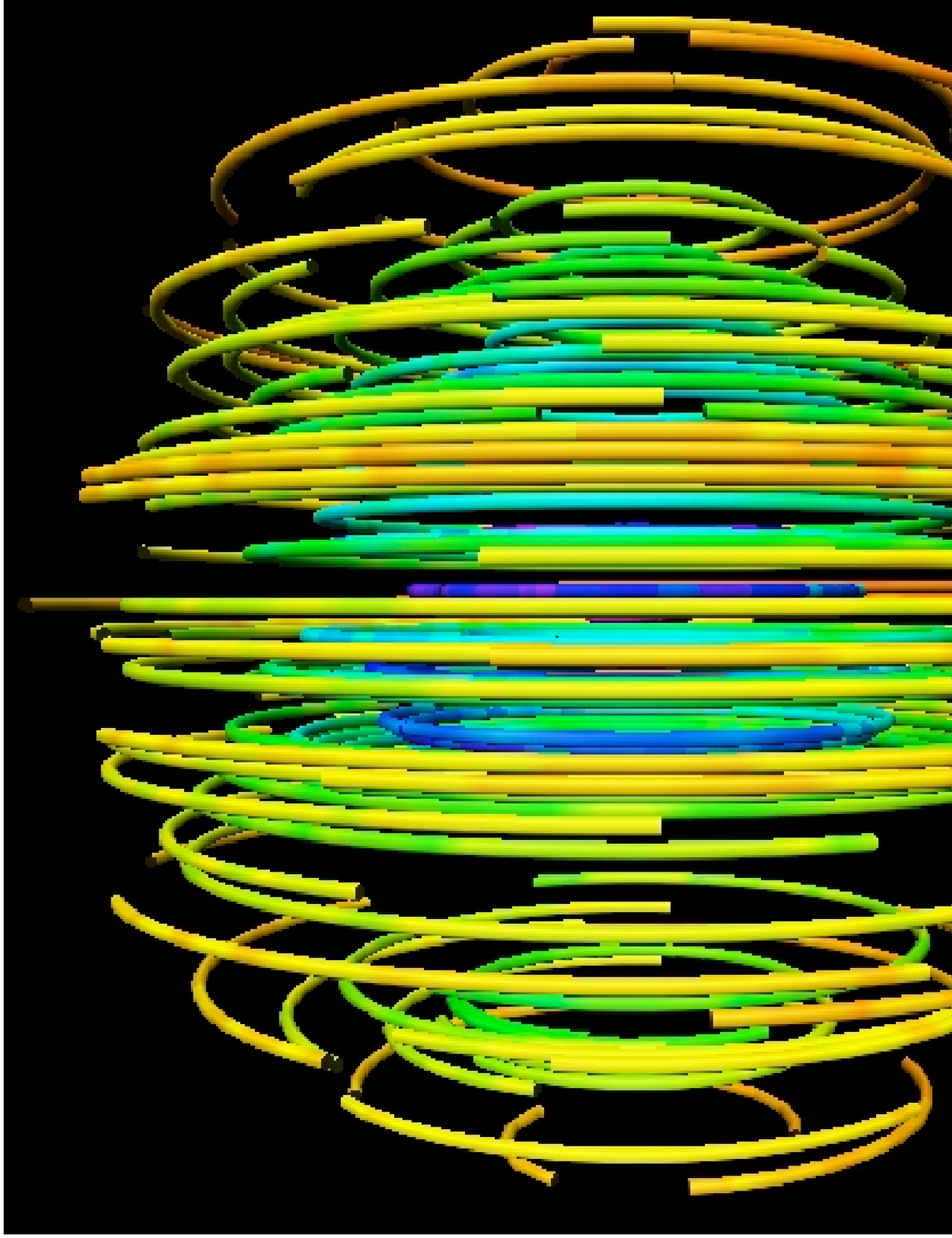}
\includegraphics[width=0.235\textwidth,angle=0]{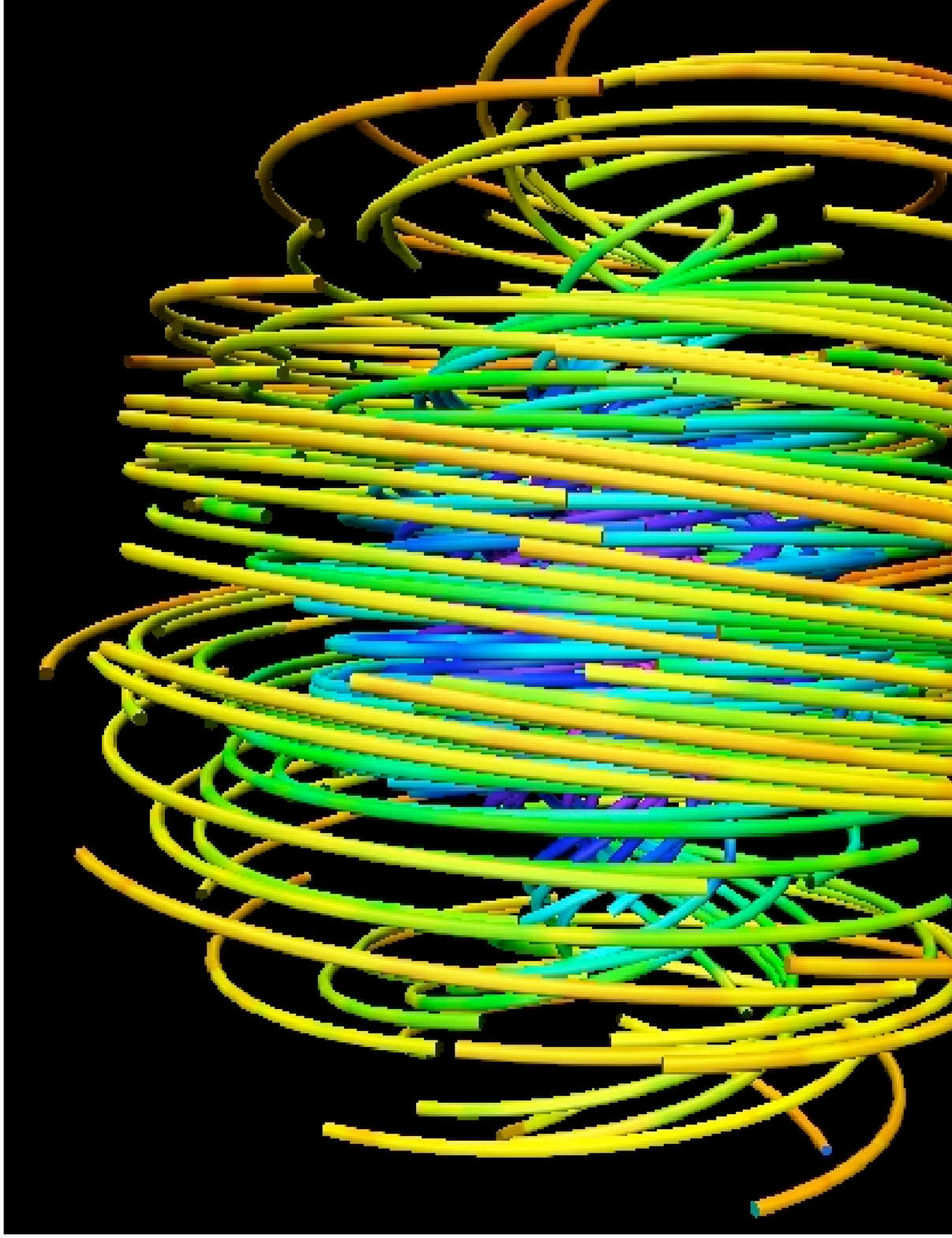}\\
\includegraphics[width=0.235\textwidth,angle=0]{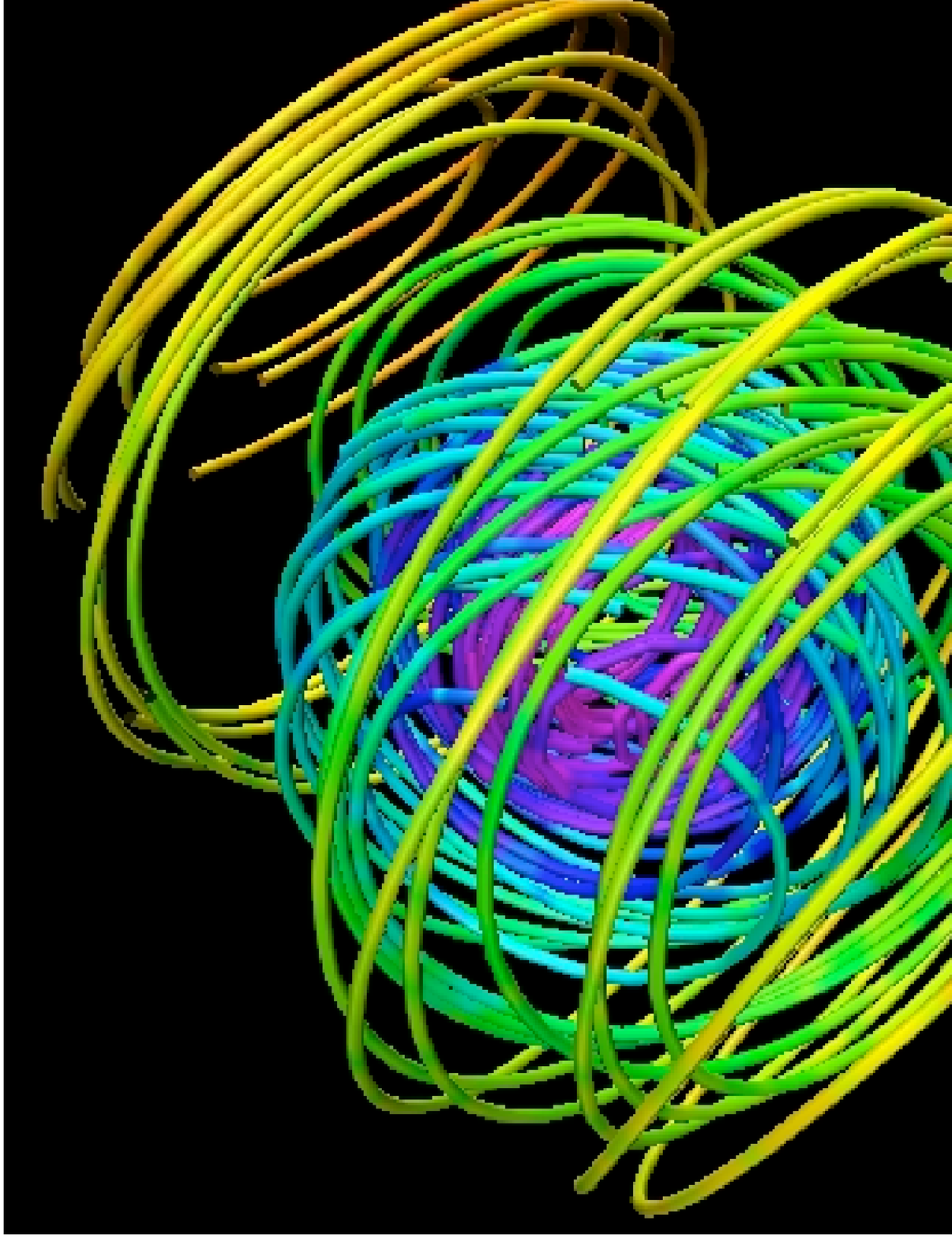}
\includegraphics[width=0.235\textwidth,angle=0]{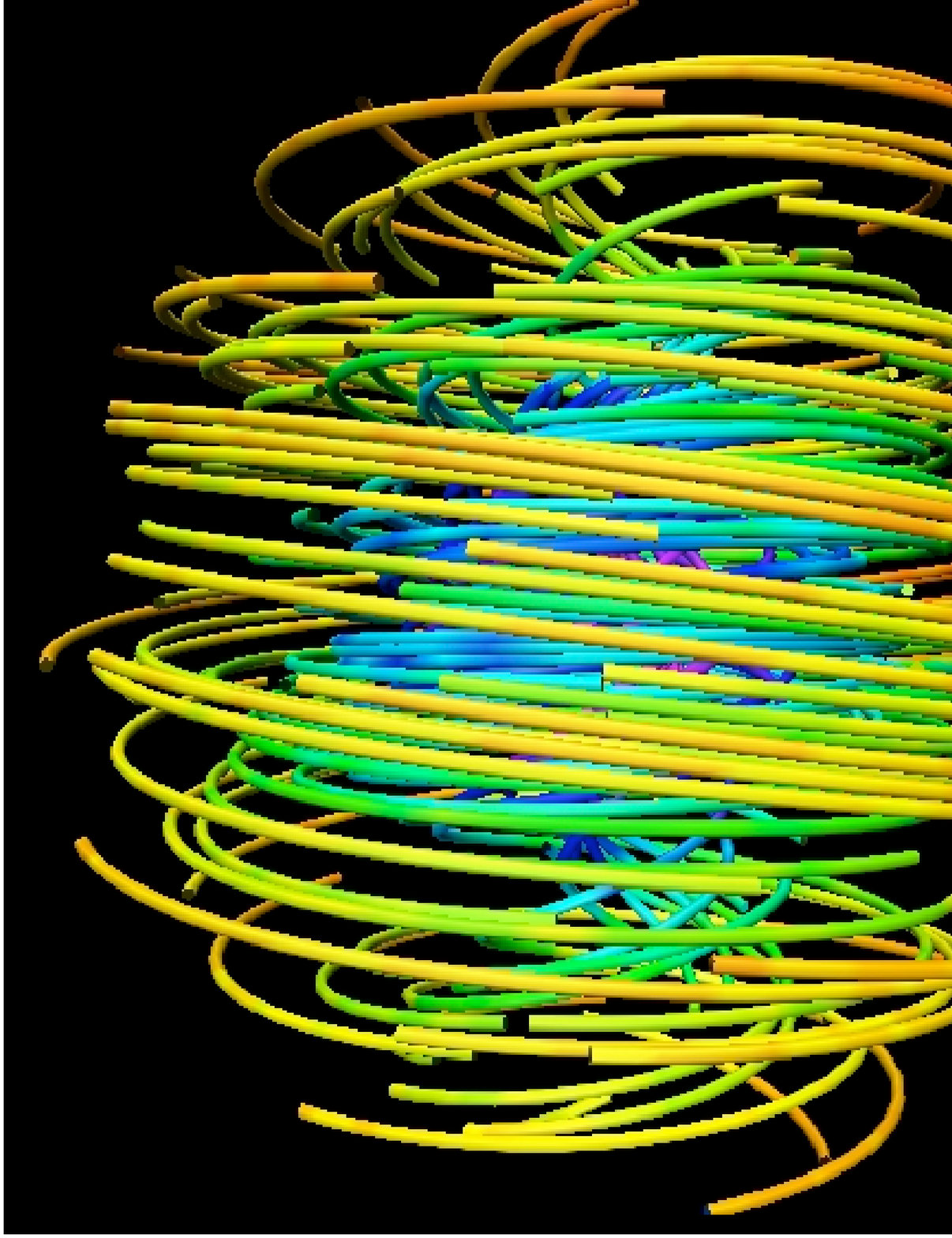}
%
\end{figure}

\begin{figure}[!htp]
\begin{center}
\includegraphics[width=0.45\textwidth]{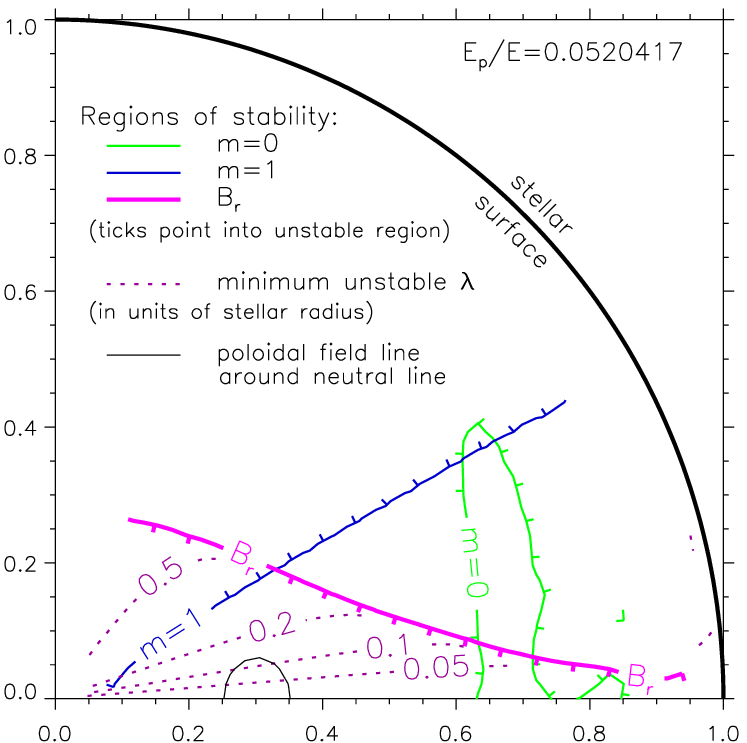}
\end{center}
\caption{Half of the meridional plane, showing the regions stable against the $m = 0$ and $1$ Tayler modes in the absence of the poloidal component, and their stabilisation by the radial component $B_r$.
\label{tayler}}
\end{figure}

\paragraph{Purely poloidal component}
The simulation is run for around ten Alfv\'en crossing times $\tau_{\rm A}$, over which time the instability grows, becomes nonlinear and results in the destruction of most of the original magnetic energy. The magnetic field amplitude is plotted at the top-left of \fig \ref{fig:mmodes}, split into components according to azimuthal wavenumber $m$; obviously at $t=0$ all the energy is in the axisymmetric $m=0$ part. The bottom-left plate of the figure shows the mean velocity in each azimuthal mode. Note the clear transition at $t\approx2\: \tau_{\rm A}$ from the linear phase to the nonlinear, reconnective phase.
\paragraph{Purely toroidal component}
The middle plates of \fig \ref{fig:mmodes} show the evolution of the toroidal field -- clearly, the $m=1$ mode is dominant.
In the lower panel of \fig \ref{fig:vapor} are drawn the magnetic field lines for the evolved configuration, i.e. at $t\approx10\: \tau_{\rm A}$. After the linear growth of the $m=1$ mode, the Tayler instability manifests itself in the nonlinear regime {\citep[cf.][]{Brun:2007a, Elstner:2008}} mainly in the movement of spherical shells relative to one another {-- which is simulated here for the first time. }
We expect eventual complete destruction of the field on a longer timescale; however a more detailed investigation is left to a forthcoming paper.
\paragraph{Mixed configuration}
The mixed poloidal-toroidal configuration exhibits completely different behaviour. The magnetic and velocity amplitudes are plotted on the right of \fig \ref{fig:mmodes}, where we see an absence of growing modes. The kinetic energy present results simply from the initial perturbation and the oscillations and waves it sets up. In \fig \ref{fig:vapor} are drawn the magnetic field lines at $t=0$ and $10\; \tau_{\rm A}$; no significant change is seen.\\ 

To better examine the potentially unstable regions, we use Tayler's stability criteria \citep{Tayler:1973} \emph{for purely toroidal fields} and estimate the stabilisation from the poloidal component, following \cite{Braithwaite:2009}. In \fig \ref{tayler} we plot Tayler's criteria for modes $m=0$ and $m=1$ -- the $m=0$ mode is unstable almost everywhere and the $m=1$ mode is unstable in a large cone around the poles; however the poloidal field stabilises these modes in most of the meridional plane except near the equatorial plane where it merely stabilises all wavelengths small enough to fit into the available space. 
{We can examine closely the behaviour of the field in the vicinity of the magnetic axis, where it can be approximated as the addition of an axial and a toroidal field (cylindrical geometry). \cite{Bonanno:2008a} outlined that in this case magnetic configurations can be subject to non-axisymmetric resonant instability. They determined the dependency of the Tayler instability maximum growth rate as a function of the azimuthal wave-number $m$ and of the ratio $\varepsilon$ of the axial field to the toroidal one. In our case, close to the center the flux function exhibits a behaviour in $\Psi \propto r^2$, so the azimuthal field is proportional to $s = r \sin \theta$ corresponding to the Bonanno et al.'s parameter $\alpha = 1$. As underlined by the authors, in that case the maximum growth rate changes remarkably slowly with $m$ for all modes with $m\geqslant2$ and the instability is weakly non-anisotropic. If we take as a value for $s_1$ the radius of the neutral line or the one where the azimuthal field is strongest, we obtain respectively $\varepsilon =0.64$ or $\varepsilon =0.79$. According to their study \citep[see][\fig 7]{Bonanno:2008a}, we fulfill the stability criterion for the modes $m=0, 1$ and $2$. Our results are therefore in agreement with their linear analysis.\\

In the simulations we run, }the mixed configuration has a poloidal energy fraction $E_{\rm p}/E=0.052$. The magnetic-to-thermal energy ratio $E/U\approx1/400$, which should mean that for stability we require $E_{\rm p}/E \gtrsim 0.04$ \citep{Braithwaite:2009}. We see then that this value of $E/U$ is near the upper limit for stability -- in other words, we are near the boundary of validity of the weak-field approximation used in Section \ref{analytic}.
%
\section{Conclusion}
Using semi-analytic methods {we derived (with an appropriate choice of boundary conditions) then tested} an axisymmetric non force-free magnetostatic equilibrium which could exist in any non-convective stellar region: the radiative core of solar-type stars, the external envelope of massive stars, and compact objects. Using numerical simulations, we {demonstrate the ability of the set-up to recover well-known instabilities in purely poloidal and toroidal cases,} then find stability of the mixed configuration under all imaginable perturbations. {We show the agreement of the result with linear analysis (limited in perturbations), highlighting the stabilizing influence of the poloidal field on the toroidal one, especially in the region close to the symmetry axis where purely toroidal fields usually develop kink-type instabilities in priority.} This is the first time the stability of an analytically-derived stellar magnetic equilibrium has been confirmed numerically.  
{This result has strong astrophysical implications: the configuration, as described in \cite{Duez:2010b}, provides a good initial condition to magneto-rotational transport to be included in next generation stellar evolution codes -- where up to now the initial field would have been chosen arbitrarily; furthermore it will help to appreciate the internal magnetic structure of neutron stars, and various astrophysical processes involving magnetars (intense activity in the X-ray and gamma-ray spectra, quasi-periodic oscillations and eventually gamma-ray bursts).}
\acknowledgements{The authors would like to thank {N. Langer, \AA. Nordlund, J.-P. Zahn, A.-S. Brun, and the {\sc MiMeS} collaboration for fruitful discussions and assistance. We are grateful to the anonymous referee for providing comments that helped in improving the manuscript. }This work was supported in part by the French PNPS (CNRS/INSU).} 
%
\bibliographystyle{aa} 

\end{document}